\documentclass[12pt]{article}

\pdfoutput=1

\usepackage{cite}

\usepackage{amssymb,latexsym}

\usepackage{epstopdf}

\usepackage{graphics,epsfig}

\usepackage{color}

\usepackage{mathrsfs}

\DeclareMathAlphabet{\mathpzc}{OT1}{pzc}{m}{it}

\usepackage{tikz}
\usetikzlibrary{arrows,shapes}
\usetikzlibrary{trees}
\usetikzlibrary{matrix,arrows} 				
\usetikzlibrary{positioning}				
\usetikzlibrary{calc,through}				
\usetikzlibrary{decorations.pathreplacing}  
\usepackage{pgffor}							

\usetikzlibrary{decorations.pathmorphing}	
\usetikzlibrary{decorations.markings}
\tikzset{
    vector/.style={decorate, decoration={snake}, draw},
	provector/.style={decorate, decoration={snake,amplitude=2.5pt}, draw},
	antivector/.style={decorate, decoration={snake,amplitude=-2.5pt}, draw},
    fermion/.style={draw=black, postaction={decorate},
        decoration={markings,mark=at position .55 with {\arrow[draw=black]{>}}}},
    fermionbar/.style={draw=black, postaction={decorate},
        decoration={markings,mark=at position .55 with {\arrow[draw=black]{<}}}},
    fermionnoarrow/.style={draw=black},
    gluon/.style={decorate, draw=black,
        decoration={coil,amplitude=4pt, segment length=5pt}},
    scalar/.style={dashed,draw=black, postaction={decorate},
        decoration={markings,mark=at position .55 with {\arrow[draw=black]{>}}}},
    scalarbar/.style={dashed,draw=black, postaction={decorate},
        decoration={markings,mark=at position .55 with {\arrow[draw=black]{<}}}},
    scalarnoarrow/.style={dashed,draw=black},
    electron/.style={draw=black, postaction={decorate},
        decoration={markings,mark=at position .55 with {\arrow[draw=black]{>}}}},
	bigvector/.style={decorate, decoration={snake,amplitude=4pt}, draw},
}

\tikzstyle{block} = [draw, rectangle, 
    minimum height=3em, minimum width=6em]

\let\a=\alpha \let\b=\beta \let\g=\gamma \let\d=\delta \let\e=\epsilon
\let\z=\zeta  \let\th=\theta  \let\k=\kappa
\let\l=\lambda \let\m=\mu \let\n=\nu \let\x=\xi \let\p=\pi 
\let\s=\sigma   \let\f=\phi  
 
      \let\G=\Gamma \let\D=\Delta \let\Th=\Theta \let\L=\Lambda
\let\X=\Xi  \let\S=\Sigma  \let\Y=\Psi
 
\let\la=\label  
  
\def\nn{\nonumber} \def\bd{\begin{document}} \def\ed{\end{document}}
\def\ds{\documentstyle} \let\fr=\frac \let\bl=\bigl \let\br=\bigr
\let\Br=\Bigr \let\Bl=\Bigl
\let\bm=\bibitem
\let\na=\nabla
\def\tU{{\widetilde U}}
\let\pa=\partial \let\ov=\overline
\def\ie{{\it i.e.\ }}
\newcommand{\be}{\begin{equation}}
\newcommand{\ee}{\end{equation}}
\def\ba{\begin{array}}
\def\ea{\end{array}}
\def\ft#1#2{{\textstyle{{\scriptstyle #1}\over {\scriptstyle #2}}}}
\def\fft#1#2{{#1 \over #2}}
\def\F#1#2{{ F_{#1}^{(#2)} }}
\def\cF#1#2{{ {\cal F}_{#1}^{(#2)} }}

\def\R{{\bf R}}
\def\sst#1{{\scriptscriptstyle #1}}
\def\oneone{\rlap 1\mkern4mu{\rm l}}
\def\e7{E_{7(+7)}}
\def\td{\tilde}
\def\wtd{\widetilde}
\def\im{{\rm i}}
\def\bog{Bogomol'nyi\ }
\newcommand{\ho}[1]{$\, ^{#1}$}
\newcommand{\hoch}[1]{$\, ^{#1}$}
\newcommand{\bea}{\begin{eqnarray}}
\newcommand{\eea}{\end{eqnarray}}
\newcommand{\ra}{\rightarrow}
\newcommand{\lra}{\longrightarrow}
\newcommand{\Lra}{\Leftrightarrow}
\newcommand{\ap}{\alpha^\prime}
\newcommand{\bp}{\tilde \beta^\prime}
\newcommand{\cB}{{\cal B}}
\newcommand{\cO}{{\cal O}}
\newcommand{\vecx}{\vec{x}}
\newcommand{\vecy}{\vec{y}}
\newcommand{\vecp}{\vec{p}}
\newcommand{\vecq}{\vec{q}}
\newcommand{\tr}{{\rm tr} }
\newcommand{\Tr}{{\rm Tr} }
\newcommand{\NP}{Nucl. Phys. }

\newcommand{\cL}{{\cal L}}
\newcommand{\cA}{{\cal A}}
\newcommand{\cT}{{\cal T}}
\newcommand{\cR}{{\cal R}}
\newcommand{\cD}{{\cal D}}
\newcommand{\cH}{{\cal H}}

\def\Cb{\bar{C}}

\def\sst#1{{\scriptscriptstyle #1}}
\def\0{{\sst{(0)}}}
\def\1{{\sst{(1)}}}
\def\2{{\sst{(2)}}}
\def\3{{\sst{(3)}}}
\def\4{{\sst{(4)}}}
\def\5{{\sst{(5)}}}
\def\6{{\sst{(6)}}}
\def\7{{\sst{(7)}}}
\def\8{{\sst{(8)}}}
\def\9{{\sst{(9)}}}
\def\p{{\sst{(p)}}}
\def\q{{\sst{(q)}}}
\def\ve{\varepsilon}
\def\vf{\varphi}
\def\F{\Phi}
\def\wg{\wedge}

\def\thb{\bar{\theta}}
\def\Thb{\bar{\Theta}}
\def\barp{\bar{p}}
\def\barq{\bar{q}}
\def\barc{\bar{c}}
\def\bard{\bar{d}}
\def\e{\epsilon}

\def \bi{\bibitem}
\def \la {\label}

\def \l {\lambda}
\def\foot{\footnote}
\def \tl  {{\tilde \l}}
\def \sql {{\sqrt \l}}
\def \adss {$AdS_5 \times S^5$\ }
\newcommand{\rf}[1]{(\ref{#1})}
\def \ov {\over}

\def\th{\theta}
\def\Th{\Theta}
\def\vth{\vartheta}
\def\btheta{{\bar\theta}}
\def\ttheta{{{\tilde\theta}}}
\def\bttheta{{{\bar\ttheta}}}
\def\vth{\vartheta}

\def\ra{\rightarrow}
\def\N{\nabla}
\def\F{{\cal F}}
\def\uM{\underline{M}}
\def\uA{\underline{A}}
\def\uN{\underline{N}}
\def\uP{\underline{P}}
\def\ua{\underline{a}}
\def\ub{\underline{b}}
\def\uc{\underline{c}}
\def\ud{\underline{d}}
\def\ue{\underline{e}}
\def\uf{\underline{f}}
\def\ui{\underline{i}}
\def\uj{\underline{j}}
\def\uk{\underline{k}}
\def\ul{\underline{l}}
\def\ual{\underline{\alpha}}
\def\ube{\underline{\beta}}
\def\um{\underline{m}}
\def\un{\underline{n}}
\def\up{\underline{p}}
\def\uq{\underline{q}}
\def\ur{\underline{r}}
\def\us{\underline{s}}
\def\umu{\underline{\mu}}
\def\unu{\underline{\nu}}
\def\ula{\underline{\l}}
\def\uka{\underline{\k}}
\def\usi{\underline{\s}}
\def\urh{\underline{\r}}
\def\cc{\circ}
\def\eqv{\equiv}

\def\ni{\noindent}

\def\Ep{E^{{}^{(+)}}}
\def\Em{E^{{}^{(-)}}}

\def\Mp{M^{{}^{(+)}}}
\def\Mm{M^{{}^{(-)}}}

\def \ha{{1\ov 2}}

\def\r{\rho}

\def\Y{{\rm Y}}
\def\X{{\rm X}}
\def\tY{\tilde{\rm Y}}
\def\tX{\tilde{\rm X}}
\def\dY{\dot{\rm Y}}
\def\dX{\dot{\rm X}}

\def \J {\mathcal{J}}
\def \del {\partial}

\def\dF{\dot{F}}
\def\dG{\dot{G}}
\def\df{\dot{f}}
\def \E {{\cal E}}
\def \S {{\cal S}}
\def \J {{\cal J}}

\def\ms{\mathcal{S}}
\def\mj{\mathcal{J}}
\def\soj{\fr{\ms}{\mj}}
\def \R {{\bf R}}
\def \om {\omega}
\def \bE {\bar E}
\def \x {{\cal X}}

\def \bi{\bibitem}
\def \la {\label}

\def \l {\lambda}
\def\foot{\footnote}
\def \tl  {{\tilde \l}}
\def \sql {{\sqrt \l}}
\def \adss {$AdS_5 \times S^5$\ }
\def \ov {\over}

\def \varpi {{\rm w}}

\def\thb{\bar{\theta}}
\def\Thb{\bar{\Theta}}
\def\mb{\bar{\m}}
\def\ab{\bar{\a}}
\def\zb{\bar{z}}
\def\psib{\bar{\psi}}
\def\barp{\bar{p}}
\def\barq{\bar{q}}
\def\barc{\bar{c}}
\def\bard{\bar{d}}
\def\e{\epsilon}
\def\wb{\bar{w}}
\def\lb{\bar{\l}}
\def\Jb{\bar{J}}
\def\Nb{\bar{N}}
\def\Zb{\bar{Z}}
\def\pab{\bar{\pa}}

\def\At{\tilde{A}}
\def\Bt{\tilde{B}}
\def\Ct{\tilde{C}}
\def\Dt{\tilde{D}}
\def\Et{\tilde{E}}
\def\Ft{\tilde{F}}
\def\Gt{\tilde{G}}
\def\Ht{\tilde{H}}
\def\Kt{\tilde{K}}
\def\Mt{\tilde{M}}
\def\Nt{\tilde{N}}
\def\Rt{\tilde{R}}
\def\at{\tilde{a}}
\def\bt{\tilde{b}}
\def\ct{\tilde{c}}
\def\dt{\tilde{d}}
\def\et{\tilde{e}}
\def\ft{\tilde{f}}
\def\htil{\tilde{h}}
\def\gt{\tilde{g}}
\def\nt{\tilde{n}}
\def\mut{\tilde{\mu}}
\def\nut{\tilde{\nu}}
\def\pht{\tilde{\f}}
\def\Pht{\tilde{\Phi}}
\def\vft{\tilde{\vf}}
\def \zet{\tilde{\z}}

\def\rht{\tilde{\rho}}

\def\asth{\hat{*}}
\def\phh{\hat{\phi}}

\def\bA{{\bf A}}

\def\ola{\overleftarrow}
\def\ora{\overrightarrow}
\def\alt{\tilde{\a}}

\def\eh{\hat{e}}
\def\eph{\hat{\e}}
\def\ph{\hat{p}}
\def\alh{\hat{\a}}
\def\beh{\hat{\b}}
\def\gah{\hat{\g}}
\def\Fh{\hat{F}}
\def\muh{\hat{\m}}
\def\nuh{\hat{\n}}
\def\thh{\hat{\th}}
\def\rhh{\hat{\r}}
\def\dh{\hat{d}}
\def\ih{\hat{i}}
\def\jh{\hat{j}}
\def\hh{\hat{h}}
\def\nh{\hat{n}}
\def\gh{\hat{g}}
\def\kh{\hat{k}}
\def\deh{\hat{\d}}
\def\wh{\hat{w}}
\def\lah{\hat{\l}}
\def\Ah{\hat{A}}
\def\Kh{\hat{K}}
\def\Nh{\hat{N}}
\def\Rh{\hat{R}}
\def\Ch{\hat{C}}
\def\Omh{\hat{\Omega}}

\def\xh{\hat{x}}

\def\ps{\rlap{\, /}\;\,p }
\def\ks{\rlap{\, /}\;\,k }

\def\gym{g_{YM}}

\def\adot{\dot{a}}
\def\bdot{\dot{b}}
\def\bpa{\bar{\pa}}

\def\pr{\prime}
\def\ssk{\medskip}
\def\clb{\color{blue}}
\def\clr{\color{red}}
\def\clg{\color{green}}

\def\bfA{{\bf A}}
\def\bfB{{\bf B}}
\def\bfK{{\bf K}}
\def\bfU{{\bf U}}
\def\bfX{{\bf X}}
\def\bfY{{\bf Y}}
\def\bfZ{{\bf Z}}
\def\bfg{{\bf g}}
\def\bfn{{\bf n}}

\def \vk{\vec{k}}
\def \vx{\vec{x}}

\begin{document}

\overfullrule=0pt
\parskip=2pt
\parindent=12pt
\headheight=0in \headsep=0in \topmargin=0in
\oddsidemargin=0in

\vspace{ -3cm}
\thispagestyle{empty}

 \vspace{0.1cm}

\setcounter{equation}{0}
\setcounter{footnote}{0}
\setcounter{section}{0}

\begin{center}

{\Large\bf Quantum gravitational effects on boundary}

\vskip 0.8cm

 \vspace{0.5cm}
 
 Frank James and I. Y. Park
 \\
 \vspace{0.3cm}
 {\it Department of Applied Mathematics,
 Philander Smith College
 \\
 Little Rock, AR 72202, USA \\
 }
 \end{center}

\begin{abstract}

Quantum gravitational effects may hold the key to some of the outstanding problems in theoretical physics.
In this work we analyze  the perturbative quantum effects on the boundary of a gravitational system and Dirichlet boundary condtion imposed at the classical level.
 Our analysis reveals that for a black hole solution there exists a clash between the quantum effects and Dirichlet boundary condition: the black hole solution of the one-particle-irreducible (1PI) action no longer obeys the Dirichlet boundary condition in the naive manner one might expect.
The analysis also suggests that the tension between the Dirichlet boundary condition and loop effects should be tied with a certain mechanism of information storage on the boundary.

\end{abstract}
\newpage

\section{Introduction}

A gravitational system is different from a nongravitational system in several peculiar ways. Among other things it displays the holographic property originating from the large amount of the diffeomorphism symmetry. With the celebrated AdS/CFT culminating in the holographic dualities, various forms of the holography have been put forth. In this work we expand on the report in \cite{Park:2016fxc} on a surprising clash between the holographic property and the widely used Dirichlet bounary condtion. The clash must also be tied with the black hole information problem \cite{Hawking:1976ra}. (See, e.g., \cite{Page:1993up,Mathur:2005zp,Polchinski:2016hrw} for reviews.)

The holographic dualities state that the boundary degrees of freedom capture the essence of the bulk physics through their couplings to the bulk modes. Let us take the prime example of AdS/CFT, the AdS$_5$/CFT$_4$ correspondence. In the original account, the decoupling limit of the coupled system of the type IIB supergravity and ${\cal N}=4$ SYM led to the duality between these two theories.   
It is possible to motivate the duality from an alternative viewpoint by starting with either one of those two theories and deducing the other \cite{Park:1999xz,Park:2001bm,Niarchos:2015moa,Grignani:2016bpq,Maxfield:2016vpw}, a view that has been reinforced by, e.g., the recent works of \cite{Sato:2002kv} and \cite{Hatefi:2012bp} in which the gauge degrees of freedom were obtained as moduli parameters of the gravity theory solutions. In that view the connection between the gauge and gravity degrees of freedom can be understood through conversions between the open and closed strings.

Recently it has been proposed in \cite{Park:2014tia}, \cite{Park:2014qoa} and \cite{Park:2015ybl} that the physical sector of a gravity theory should be described by a hypersurface at the boundary. This holographic property (or ``duality") may appear different from that of the typical AdS/CFT-type dualities in that  in the latter the gauge theory degrees of freedom may seem alien to the original gravity system whereas in the former the degrees of freedom ``dual" to the original gravity theory are akin to the gravity theory.\footnote{In this sense, the present work and  \cite{Park:2016fxc}  are in line with the pre-AdS/CFT works \cite{Benguria:1976in,Witten:1988hf,Balachandran:1991dw,Smolin:1995vq} in which the boundary dynamics were studied.} But this is not true: in both cases the dual holographic (i.e., lower dimensional) degrees of freedom are part of the original gravity theory.	The real difference between AdS/CFT type dualities and the duality proposed in \cite{Park:2014tia} is the manner in which the dual degrees of freedom are obtained: in the case of \cite{Park:2014tia} they are directly visible in the sense that they do not require any transformation to become recognizable whereas the gauge degrees of freedom in AdS/CFT become recognizable after a certain ``dualization process" involving the Hamilton-Jacobi formalism \cite{Sato:2002kv}\cite{Hatefi:2012bp}.

The equations of motion of a field theoretic system are partial differential equations and one must impose boundary conditions in order to find a solution. Two very commonly used boundary conditions are Dirichlet and Neumann conditions.  In string theory, the physics of the Neumann boundary condition was developed before the Dirichlet boundary condition; in quantum field theory and gravity, the Dirichlet boundary condtion has been widely used. (Recent discussions on the boundary conditions include \cite{Krishnan:2016mcj} and \cite{Lehner:2016vdi}.)
In light of the proposal  \cite{Park:2014tia,Park:2015ybl} one is naturally led to a question of whether or not the holographic property would be compatible with the Dirichlet boundary condition. Evidently, the presence of the active degrees of freedom at the boundary is at odds with what's required by the Dirichlet boundary condition: the former implies dynamical boundary fields whereas the latter requires the fields dying out  and not varying at the boundary. As a matter of fact, a new approach to the boundary dynamics has recently been adopted in the framework of loop quantum gravity, and the theory of the boundary dynamics has been identified in a manner more concrete than ever \cite{Freidel:2016bxd}\cite{Donnelly:2016auv}. Notably, it was proposed that the entire set of the boundary conditions will be required for the complete description of the Hilbert space of the gravity theory.

 An explicit demonstration of the `violation' of the Dirichlet boundary condition involves highly technical steps for several reasons.
Firstly, the demonstration requires analysis of a time- and position- dependent solution at the classical level to start with (and later at the quantum level as well). Finding such a solution is a complicated task even at the classical level. We will, following the literature \cite{Murata:2010dx}, rely on a series form of the solution.
Secondly, computation of the one-particle-irreducible (1PI) effective action - which must precede the computation of the quantum-corrected equations of motion - is complicated for a time- and position- dependent background. Obtaining merely the Green's functions would be nontrivial.

Fortunately, however, some of these complications can be avoided because they are inessential to our goal of observing the violation of the Dirichlet boundary condition. For our goal one can focus on the finite pieces
contained within the divergences. The divergent parts are common to any background since they are expected to arise from a flat limit. 
For precise perturbative computation of the 1PI action it would be appropriate\footnote{The choice of the background would not matter if one could consider all of the nonperturbative contributions as well. Within the perturbative calculation, the proper way would be to expand around the background of interest.} to expand the action around the background of interest (the time- and position- dependent one in the present case) and obtain the Green's functions and  1PI action associated with that background. Instead, we consider the vacuum, i.e., an AdS spacetime; we expand the classical action around it and obtain the one-loop 1PI action. To recapitulate, expanding around the AdS background cannot be entirely justified as a step to obtain the perturbative 1PI action from which the quantum effects on the boundary of the time- and position- dependent solution is studied. Again, however, we will focus on the divergences and subsequently the finite pieces therein contained for which the procedure  is justified (since the divergences must be common to any background) and thus the present procedure will serve the purpose of revealing the effects that the quantum corrections will have on the boundary.

As we will show below the quantum corrections induce time-dependent boundary modes, thereby, leading to violation of the Dirichlet boundary condition. The subsequent analysis also reveals that the violation cannot be re-interpreted as a Neumann boundary condition. It may indicate the existence of the boundary's own full dynamics. (More on this in relation to \cite{Freidel:2016bxd}\cite{Donnelly:2016auv}.)
The violation of Dirichlet boundary condition at the quantum level - which we believe should generically occur to a gravitational system - should have an important implication for the black hole information paradox. 
A certain mechanism of the information storage on the boundary must be responsible for the violation.
As we will see, the bulk modes partially depend on the boundary modes, a feature not shared by the classical analysis due to the Dirichlet boundary condition adopted.

\vspace{.2in}

The rest of the paper is organized as follows:

In section 2 we begin by reviewing several ingredients to set the stage for the main analysis in the subsequent sections.
In section 3 we take a 4D AdS gravity-scalar system and enumerate various relevant loop diagrams and obtain the counter-terms. Throughout, we adopt the conventional framework and limit the loop analysis to one-loop. We employ dimensional regularization. In section 4 we illustrate the idea of the violation of the Dirichlet boundary condition with the time- and position- dependent solution reviewed in section 2 but this time extended to the quantum level. We check the renormalizability of the system by outlining the renormalization procedure by employing the generalized modified-minimal-subtraction scheme for the finite parts. The solution receives quantum corrections, and as we will see, the quantum-corrected solution does indeed violate the Dirichlet boundary condition imposed at the classical level. We comment on the mechanism of the information storage on the boundary.
In section 5 we conclude with a summary and future directions. Our conventions are listed in Appendix A; some details of the computations can be found in Appendix B.

\section{Setup}

In this section we set the ground for the main task in section 3 and 4 where we analyse the effects of the loop corrections on the classical solution with the Dirichlet boundary condition. The system we consider is the following gravity-scalar system \cite{Gubser:2008px}\cite{Hartnoll:2008vx}\cite{Murata:2010dx}\cite{Bhaseen:2012gg}:
\begin{equation}
S=\int d^3 x \sqrt{-g}\left[R+\frac{6}{L^2}-\fr12(\pa_\mu \z)^2 -\fr12m^2\z^2 \right]  \la{sys}
\end{equation}
where $m^2=-\fr{2}{L^2}$.

\subsection{expansion and Green's functions}

Let us start with the gravity sector (the cosmological constant terms have been set apart for now):  
\bea
S_g=\fr1{\k^2}\int d^4 x \sqrt{-g}\;R  
\la{unsplit}
\eea
By shifting the metric around the unperturbed metric $g_{0\m\n}$ \cite{Antoniadis:1995fc}\cite{Park:2015ota}
\bea
g_{\m\n}\equiv  h_{\m\n}+\tilde{g}_{{}_B\m\n}\quad \mbox{where}\quad \tilde{g}_{{}_B\m\n}\equiv \vf_{{}_B\m\n}+g_{0\m\n} \la{gshift}
\eea
where $\vf_{{}_B\m\n}$ and $h_{\m\n}$ denote the background field and fluctuation respectively one obtains\footnote{The overall factor $\fr1{\k'^2}$ where $\k'^2 \equiv 2\k^2 $ has been suppressed.}
\bea
\hspace{1in} &&\cL_g = \sqrt{-\gt}\,\Big( -\fr12\tilde{\N}_\g h^{\a\b}\tilde{\N}^\g h_{\a\b}+\fr14 \tilde{\N}_\g h^{\a}_\a \tilde{\N}^\g h^{\b}_\b  \la{gravcubcov}
\\
&&\hspace{-1in}+h_{\a\b}h_{\g\d}\Rt^{\a\g\b\d}-h_{\a\b}h^{\b}{}_\g \Rt^{\k\a\g}{}_{\k}
+h^{\a}{}_{\a}h_{\b\g}\Rt^{\b\g}-\fr12 h^{\a\b}h_{\a\b}\Rt
+\fr14  h^{\a}_\a  h^{\b}_\b \Rt +\cdots\Big) \nn
\eea
where $(\cdots)$ denotes the terms relevant only for higher-loops. The objects with the tildes indicate that they are the background quantities constructed out of $\tilde{g}_{{}_B\m\n}$. The scalar sector expansion is given by
\bea
&&\hspace{.8in}\cL_\z= -\fr12\sqrt{-g}\; g^{\m\n}(\pa_\m\z) (\pa_\n \z) -\fr{m^2}2\sqrt{-g}\;  \z^2 \\
&&  \hspace{-.4in}= -\fr12\sqrt{-\gt_B}\Big[\gt_B^{\m\n}-h^{\m\n}+\fr12 h\gt_B^{\m\n}+h^{\m\r}h_{\r}^\n+\fr18 \gt_B^{\m\n}(h^2-2h_{\r\s}h^{\r\s})-\fr12 hh^{\m\n}+\cdots \Big] (\pa_\m\z)  (\pa_\n \z)         \nn\\
&&  \hspace{.8in}-\fr{m^2}2\sqrt{-\gt_B}\Big[1+\fr12 h+\fr18 (h^2-2h_{\r\s}h^{\r\s})+\cdots \Big]  \z^2  \la{sgsector}
\eea
Below a shift in the scalar field, $\z\rightarrow \z_B+{\z}$ will also be considered.

\subsubsection*{scalar case}

Let us consider an AdS background in the Poincare coordinates:
\bea
ds^2=\fr1{z^2}(dz^2+(dx^m)^2)
\eea
and the Green's function (or the Feynman propagator) in that background. 
The Green's function for the scalar part satisfies
\bea
(\nabla_{0\m}\nabla_0^\m-m^2)G_s=\fr{i}{\sqrt{-g}}  \d (x-y)
\eea
The Green function is known in a closed form \cite{Burgess:1984ti,Inami:1985wu,D'Hoker:2002aw,D'Hoker:1999jc}
\bea
G_s=i\fr{2^{-\D}C_\D}{2\D-d}\;\xi^\D\;F\Big(\fr{\D}2, \fr{\D}2+\fr12;\D-\fr{d}2+1;\xi^2\Big)
\eea 
where $F$ is the hypergeometric function and
\bea
\xi=\fr{2z_0w_0}{z_0^2+w_0^2+(\vec{z}-\vec{w})^2}\quad,\quad C_\D=\fr{\G(\D)}{\pi^{d/2}\G(\D-d/2)} \la{Cdef}
\eea
The parameter $\Delta$ is related to the mass by $m^2=\D(\D-d)$, i.e.,
\bea
\D_\pm=\fr{d}2\pm \sqrt{\fr{d^2}4+m^2}
\eea
For the 4D scalar case under consideration one has
\bea
d=3\quad,\quad \D=1 \quad,\quad m^2=-2
\eea
The regularized value of $C_1$ is $-\fr12 \pi^2$.

\subsubsection*{graviton case}

The graviton propagator associated with the traceless fluctuation mode \cite{Park:2015ota}\cite{Park:2015xoa} is given by
\bea 
<h_{\m\n}(x_1)h_{\r\s}(x_2)>&=& P_{\m\n\r\s}\, \D(x_1-x_2) 
\eea
where $\D(x_1-x_2)$ is given by the Green's function in the scalar case: 
\bea
\D(x_1-x_2)=G_s(x_1-x_2);
\eea
the tensor $P_{\m\n\r\s}$ is given by
\bea
P_{\m\n\r\s} &\equiv& \fr12\Big(\gt_{B\m\r}\gt_{B\n\s}+\gt_{B\m\s}\gt_{B\n\r}
- \fr12\gt_{B\m\n}\gt_{B\r\s}\Big)  \nn\\ 
&\simeq &\fr12\Big(g_{0\m\r}g_{0\n\s}+g_{0\m\s}g_{0\n\r}
- \fr12g_{0\m\n}g_{0\r\s}\Big)
\eea
where the righ-hand side of the second equality is the leading order $\vf_{B\m\n}$-expansion of $P_{\m\n\r\s}$.
Note that $P_{\m\n\r\s}$ is traceless in $(\m\n)$- and $(\r\s)$- indices. (As a matter of fact, the necessity of employing a traceless propagator 
was explicitly stated earlier in \cite{Ortin} as we have recently become aware.)

\subsection{review of the classical solution}

Let us consider the following metric ansatz \cite{Murata:2010dx} for the solution of the system \rf{sys}:
\begin{eqnarray}
ds^2&=&-\frac{1}{z^2}\left[F(t,z) dt^2 + 2 dt\,dz\right] + \Phi(t,z)^2 (dx^2+dy^2)\nonumber\\
\z&=&\z(t,z)\nonumber
\end{eqnarray}
with
\bea
F(t,z)&=&F_0(t) +F_1(t) z+ F_2(t)z^2+F_3(t)z^3 + ...\nonumber\\
\Phi(t,z)&=&\frac{1}{z}+\Phi_0(t) +\Phi_1(t) z+ \Phi_2(t)z^2+\Phi_3(t)z^3 + ...\nonumber\\
\z(t,z)&=&\z_0(t) +\z_1(t) z+ \z_2(t)z^2+\z_3(t)z^3 + ...\la{clasol}
\eea
As observed in \cite{Murata:2010dx}, there is a residual gauge symmetry: $\fr1{z}\ra \fr1{z}+g(t)$. Because of this, the component $ \Phi_0$ can be gauge-fixed, and the classical bulk geometry is the same regardless of the choice. (More on this later.)
Once one sets $ \Phi_0=0$ - which is consistent with the Dirichlet boundary condition - and substitutes these ansatze into the field equations, one gets, for the first several modes,
\bea
&&\z _0=0,\quad F_0=1, \quad F_1=0, \quad \Phi _1=-\frac{1}{8} \z _1^2, \quad F_2=-\frac{1}{4} \z _1^2,\quad \Phi _2=-\frac{1}{6} \z _1 \z_2\nn\\
&& \hspace{-.5in} \z _3=\frac{1}{2} \left(\fr12 \z _1{}^2 {\z}_1+2 \dot{\z} _2\right),\quad \Phi _3=\fr1{96}\Big[ -\fr{11}4 \z _1{}^4 -8 \z _2^2 -12 {\z}_1 \dot{\z} _2\Big],\quad \dot{F}_3=-\fr12\z_1\dot{\z}_2+\fr12 \z_1\ddot{\z}_2\nn\\
\eea
 The results $\z _0=0,F_0=1$ are also consistent with the Dirichlet boundary conditions imposed on the corresponding fields. As we will see in section 4, it is the similar quantum level consistency that will be compromised once the perturbative loop effects are taken into account, leading to the violation of the Dirichlet condition.

\section{One-loop computation}

To obtain the quantum-corrected field equations one should first compute various one-loop diagrams and obtain the one-loop 1PI action by using the propagators reviewed in the previous section. The quantum-corrected field equations will follow by taking a variation of the 1PI action in the standard manner. The computations in this section are carried out entirely within the conventional framework \cite{Buchbinder} (other than employing the traceless graviton propagator \cite{Park:2015ota}\cite{Park:2015xoa}\footnote{As mentioned previously, the necessity of the traceless condition, $h=0$, had been noted earlier in the literature (e.g. in \cite{Ortin}).}).

We employ dimensional regularization and focus on the divergent parts coming with the gamma function $\G(\ve)$, $\ve\equiv 2-\fr{D}{2}$. The gamma function can be expanded 
\bea
\G(\ve)=\fr1{\ve}-\g+\cdots \la{ge}
\eea
where  $\g$ denotes the Euler constant.
Although the renormalization removes the divergent parts the remaining finite terms are arbitrary: they get fixed by a renormalization scheme.
Our focus will be on those finite parts. For instance, the minimal subtraction (MS) scheme removes the $\fr1{\ve}$ term and leaves the finite part in \rf{ge}. 
The modified MS scheme, $\overline{\mbox{MS}}$, is popular in dimensional regularization and it removes the $\fr1{\ve}$ term with additional finite parts (such as $\g$ and $\log 4\pi$). We will employ the {\em generalized} $\overline{\mbox{MS}}$ scheme and keep the finite parts unspecified.

To compute the 1PI action perturbatively it would be appropriate to expand the action around the background of interest and obtain the Green's function and  1PI action associated with that background. However, because this task is simply too complicated one may take a `shortcut' approach: one may consider the vacuum, i.e., an AdS spacetime and expand the classical action around it and work out the 1PI action. 
Since the divergences must come from the flat limit and one seeks after the divergences associated with $\G(\ve)$ (and eventually the finite parts contained within $\G(\ve)$), one may consider the flat space limit of the AdS and examine the leading behaviors (see Appendix B for details). 
At the end the analysis produces the results identical to those that would be obtained by applying the background field method (BFM) entirely in the flat spacetime from the beginning.
We illustrate this with one of the diagrams below (Fig.4 (a)) and carry out the rest of the divergence analyses of the other diagrams in a flat background.


In the loop expansion all of the one-loop terms come with the same power of the Planck's constant $\hbar$, which we keep implicit. Some of the diagrams considered will have higher powers of the Newton's constant. We consider them to show that the renormalization program can be extended to the subleading orders in $\k$. Since our focus is on the divergences we consider the flat background for the diagrams below. (The illustration that the AdS background computation yields the same result as the flat background computation can be found in Appendix B.)

\subsection{scalar-involving sectors}

\subsubsection*{diagrams with scalar loop}

Let us consider the diagrams with a scalar loop. 
There are three diagrams with two external graviton legs:
\bea
&&\hspace{-.4in}
\begin{tikzpicture}[line width=1 pt, scale=1]
\draw[vector] (0:-1)--(0,0);
\draw (.52,0) circle (.49cm);	

\node at (-22.1:1.1) {$\z$};
\node at (20.1:1.15) {$\z$};

\node at (-80:.5) {$\z$};
\node at (80:.5) {$\z$};

\begin{scope}[shift={(1,0)}]
\draw[vector] (0:1)--(0,0);
\node at (0:1.4) {$\vf_B$};
\node at (0:-2.4) {$\vf_B$};
\node at (70:-1.2) {(a)};	
\end{scope}
\end{tikzpicture}
\begin{tikzpicture}[line width=1 pt, scale=1]
\draw[vector] (0:-1)--(0,0);
\draw (.52,0) circle (.49cm);	

\node at (-22:1.2) {$\pa\z$};
\node at (21.5:1.2) {$\pa\z$};

\node at (-80:.5) {$\z$};
\node at (80:.6) {$\z$};

\begin{scope}[shift={(1,0)}]
\draw[vector] (0:1)--(0,0);
\node at (0:1.4) {$\vf_B$};
\node at (0:-2.4) {$\vf_B$};
\node at (70:-1.2) {(b)};	
\end{scope}
\end{tikzpicture} 
\begin{tikzpicture}[line width=1 pt, scale=1]
\draw[vector] (0:-1)--(0,0);
\draw (.52,0) circle (.49cm);	

\node at (-25:1.2) {$\pa\z$};
\node at (25.5:1.2) {$\pa\z$};

\node at (-77:.6) {$\pa\z$};
\node at (80:.6) {$\pa\z$};

\begin{scope}[shift={(1,0)}]
\draw[vector] (0:1)--(0,0);
\node at (0:1.4) {$\vf_B$};
\node at (0:-2.4) {$\vf_B$};
\node at (70:-1.2) {(c)};	
\end{scope}
\end{tikzpicture} \nn\\
&&\hspace{1in} \mbox{Figure 1: diagrams with scalar loop}  \nn
\eea
One of the two relevant vertices in the diagrams  in Fig. 1 comes from
\bea
&& -\fr12\sqrt{-g}\; g^{\m\n}(\pa_\m\z) (\pa_\n \z) \rightarrow -\fr12\sqrt{-\gt_B}\Big[\gt_B^{\m\n}-h^{\m\n}+\fr12 h\gt_B^{\m\n}\nn\\
&& +h^{\m\r}h_{\r}^\n+\fr18 \gt_B^{\m\n}(h^2-2h_{\r\s}h^{\r\s})-\fr12 hh^{\m\n}+\cdots \Big] (\pa_\m { \z})  (\pa_\n { \z})      
\eea
Among these the relevant term is
\bea
&& -\fr12\sqrt{-\gt_B}\;\gt_B^{\m\n}  (\pa_\m{ \z})  (\pa_\n { \z}) \ra   V_{\vf_B\pa\z\pa\z}\equiv
\fr12\;\vf_B^{\m\n}  (\pa_\m{ \z})  (\pa_\n { \z})
\la{sgsector}  
\eea
The other vertex comes from the mass term:
\bea
-\fr{m^2}2\sqrt{-g}\;  \z^2 =
-\fr{m^2}2\sqrt{-\gt_B}\Big[1+\fr12 h+\fr18 (h^2-2h_{\r\s}h^{\r\s})+\cdots \Big]  { \z}^2     
\eea
The relevant vertex comes from
\bea
& &
{-}\fr{m^2}{2}\sqrt{-\gt_B}\; \z^2 \ra  V_{\vf_B\z \z}\equiv  { -}\fr{m^2}{ 4} \vf_B\; \z^2\
\la{sgsectortwo}  
\eea
However, this vertex does not play a role because of the gauge-fixing of the trace mode, $\vf_B$ \cite{Park:2015ota}.
The first two diagrams vanish for the same reason: the external graviton lines contain the vertex $V_{\vf_B\z \z}$ and the trace of the fluctuation metric $\vf_B$ has been gauge-fixed to zero. The correlator for Fig.1 (c) is given by
\bea
-\fr12<\int V_{\vf_B\pa\z\pa\z}\; \int V_{\vf_B\pa\z\pa\z}>
=-\fr18\int\int \vf_B^{\r\s}\vf_B^{\r'\s'}<(\pa_\r\z \pa_\s\z)(\pa_{\r'}\z\pa_{\s'}\z)>\nn\\
\eea
which leads, in, e.g., the $\overline{MS}$ scheme, to the following one-loop counter-term: 
\bea
\D \cL= \fr{\G(\ve)}{(4\pi)^2}  \fr1{120} \Big(\fr12 \Rt^2+\Rt_{\a\b}\Rt^{\a\b}\Big)
\eea

\subsubsection*{diagrams with both scalar and graviton legs}

Let us consider the diagrams with with both the graviton and scalar lines:
\bea
&& \hspace{-.3in}
\begin{tikzpicture}[line width=1 pt, scale=1]
\draw (-140:1)--(0,0);
\draw (140:1)--(0,0);
\draw[vector] (.52,0) circle (.49cm);	

\node at (-140:1.4) {$\pa\z_B$};
\node at (140:1.4) {$\pa\z_B$};

\begin{scope}[shift={(1,0)}]
\draw[vector] (0:1)--(0,0);
\node at (0:1.4) {$\gt_B$};
\node at (70:-1.2) {(a)};	
\end{scope}
\end{tikzpicture} 
\hspace{.8in}
\begin{tikzpicture}[line width=1 pt, scale=1]
\draw (-140:1)--(0,0);
\draw (140:1)--(0,0);
\draw[vector] (.52,0) circle (.49cm);	

\node at (-140:1.2) {$\z_B$};
\node at (140:1.2) {$\z_B$};

\begin{scope}[shift={(1,0)}]
\draw[vector] (0:1)--(0,0);
\node at (0:1.3) {$\gt_B$};
\node at (70:-1.2) {(b)};	
\end{scope}
\end{tikzpicture} \nn\\
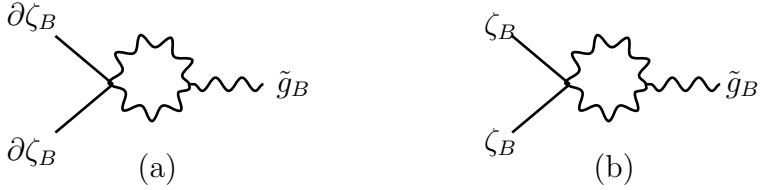
&&\hspace{-.2in} \mbox{Figure 2: diagrams with scalar and graviton lines}  \nn
\eea
With a shift in the scalar field, $\z\rightarrow \z_B+{\z}$, one of the relevant vertices comes from
\bea
&& -\fr12\sqrt{-g}\; g^{\m\n}(\pa_\m\z) (\pa_\n \z) \rightarrow -\fr12\sqrt{-\gt_B}\Big[\gt_B^{\m\n}-h^{\m\n}+\fr12 h\gt_B^{\m\n}\nn\\
&& +h^{\m\r}h_{\r}^\n+\fr18 \gt_B^{\m\n}(h^2-2h_{\r\s}h^{\r\s})-\fr12 hh^{\m\n}+\cdots \Big] (\pa_\m\z_B)  (\pa_\n \z_B)      
\eea
Among these the relevant terms are
\bea
&& V_{hh\pa\z_B\pa\z_B}\equiv -\fr12\sqrt{-{g_0}}\Big[h^{\m\r}h_{\r}^\n-\fr14 g_0^{\m\n}(h_{\r\s}h^{\r\s}) \Big] (\pa_\m\z_B)  (\pa_\n \z_B) 
\la{sgsector}  
\eea
Similarly the relevant vertex of the expansion
\bea
-\fr{m^2}2\sqrt{-g}\;  \z^2 \rightarrow 
-\fr{m^2}2\sqrt{-\gt_B}\Big[1+\fr12 h+\fr18 (h^2-2h_{\r\s}h^{\r\s})+\cdots \Big]  \z_B^2     
\eea
is
\bea
& &V_{hh\z_B \z_B}\equiv
\fr{m^2}8\sqrt{-g_0}\;h_{\r\s}h^{\r\s}  \z_B^2
\la{sgsectortwo}  
\eea
By expanding \rf{gravcubcov} one gets
\bea
\cL&=& -\fr12 {\pa}_\g h^{\a\b}{\pa}^\g h_{\a\b}+\fr14 {\pa}_\g h^{\a}_\a {\pa}^\g h^{\b}_\b  \nn\\
&& + \cL_{V_I}+\cL_{V_{II}}+\cL_{V_{III}}   \la{eawv}
\eea
where
\bea
\cL_{V_I} &=&   \Big(2\eta^{\b\b'}\tilde{\G}^{\a' \g\a}- \eta^{\a\b}\tilde{\G}^{\a' \g\b'}\Big)\pa_\g h_{\a\b}\, h_{\a'\b'}  \nn\\
\cL_{V_{II}} &=& \Big[\fr12(\eta^{\a\a'}\eta^{\b\b'}\vf^{\g\g'}+\eta^{\b\b'}\eta^{\g\g'}\vf^{\a\a'}
+\eta^{\a\a'}\eta^{\g\g'}\vf^{\b\b'})\nn\\
&&-\fr14 \vf\, \eta^{\a\a'}\eta^{\b\b'}\eta^{\g\g'}-\fr12 \eta^{\g\g'}\eta^{\a'\b'}\vf^{\a\b}  \nn\\
&&+\fr14 (-\vf^{\g\g'}+\fr12 \vf \eta^{\g\g'})\eta^{\a\b}\eta^{\a'\b'}
\Big] \pa_\g h_{\a\b}\, \pa_{\g'}h_{\a'\b'}  \la{lv12}
\eea
\bea
\hspace{-.2in}\cL_{V_{III}} = \sqrt{-\gt}\Big( h_{\a\b}h_{\g\d}\Rt^{\a\g\b\d}-h_{\a\b}h^{\b}{}_\g \Rt^{\k\a\g}{}_{\k}
+h^{\a}{}_{\a}h_{\b\g}\Rt^{\b\g}-\fr12 h^{\a\b}h_{\a\b}\Rt
+\fr14  h^{\a}_\a  h^{\b}_\b \Rt \Big)  \la{gver}  \nn\\
\eea
These two vertices come from the first line of \rf{gravcubcov}. (The distinction between $\cL_{V_I} $ and $\cL_{V_{II}}$ is for convenience in Mathematica coding.) The vertex $\cL_{V_{III}}$ is just the third line of \rf{gravcubcov}
The diagram in Fig.2 (a) yields a vanishing result and thus need not be further considered.
As for the diagram in Fig.2 (b), one should evaluate
\bea
&&   \fr{i^2}{2} \Big<2 \int(\cL_{V_{I}}(x)+\cL_{V_{II}}(x)+\cL_{V_{III}}(x)) \;\int V_{hh\z_B\z_B}(x') \Big> \nn\\ 
&=& \fr{{ m^2}}{16} \int\int \z_B^2\Big< (\cL_{V_{I}}+\cL_{V_{II}}+\cL_{V_{III}})  (h^2-2h_{\m\n} h^{\m\n}) \Big>
\eea
The result turns out to be
\bea
\D \cL=\k'^2 \fr{{ m^2}}{16} \fr{\G(\ve)}{(4\pi)^2}(-9\Rt)\z_B^2=- \k'^2 \fr{\G(\ve)}{(4\pi)^2} \fr{9 { m^2}}{16} \;\z_B^2\,\Rt
\eea
Note that this result is higher in $\k^2$ than the pure gravity diagrams below; it will be relevant when we consider renormalization at the first few subleading orders of $\k$ in section 4.

\subsection{pure gravity sector}

As far as the graviton external legs are concerned we limit our analysis to the diagrams with up to (and including) two graviton legs.
There are two diagrams:
\bea
&&
\begin{tikzpicture}[line width=1 pt, scale=1]
\draw[vector] (0:-1)--(0,0);
\draw[vector] (.52,0) circle (.49cm);	



\begin{scope}[shift={(1,0)}]
\draw[vector] (0:1)--(0,0);
\node at (0:1.4) {$\vf_B$};
\node at (0:-2.4) {$\vf_B$};
\node at (70:-1.4) {(a) graviton loop };
\end{scope}
\end{tikzpicture} 
\hspace{.2in}
\begin{tikzpicture}[line width=1 pt, scale=1]
\draw[vector] (0:-1)--(0,0);
\draw[dashed] (.52,0) circle (.49cm);	



\begin{scope}[shift={(1,0)}]
\draw[vector] (0:1)--(0,0);
\node at (0:1.4) {$\vf_B$};
\node at (0:-2.4) {$\vf_B$};
\node at (70:-1.4) {(b) ghost loop};
\end{scope}
\end{tikzpicture} \nn\\
&&\hspace{1in}
\mbox{Figure 3   gravity sector}\nn
\eea
After a lengthy algebra one can show that the total one-loop counterterm in, e.g., the $\overline{MS}$ scheme is given by \cite{Park:2015ota}
\bea
\D \cL
&=& \fr12\fr{\G(\e)}{(4\pi)^2}\Big[\fr{41}{60}\Rt_{\m\n}\Rt^{\m\n}{ +\fr{27}{40}}\Rt^2\Big]
\eea

\subsection{diagrams with four external scalar lines}

There are three diagrams coming with $\k^4$ power:
\bea
&& \hspace{-.2in}
\begin{tikzpicture}[line width=1 pt, scale=1]
\draw (-140:1)--(0,0);
\draw (140:1)--(0,0);
\draw[vector] (.52,0) circle (.47cm);	

\node at (-137:1.3) {$\pa\z_B$};
\node at (137:1.3) {$\pa\z_B$};

\begin{scope}[shift={(1,0)}]
\draw (-40:1)--(0,0);
\draw (40:1)--(0,0);
\node at (-38:1.3) {$\pa\z_B$};
\node at (38:1.3) {$\pa\z_B$};	
\node at (65:-1.2) {(a)};	
\end{scope}
\end{tikzpicture}
\hspace{.3in}
\begin{tikzpicture}[line width=1 pt, scale=1]
\begin{scope}[shift={(1,0)}]
\draw (-140:1)--(0,0);
\draw (140:1)--(0,0);
\draw[vector] (.52,0) circle (.47cm);	

\node at (-140:1.2) {$\z_B$};
\node at (140:1.2) {$\z_B$};

\draw (-21:2)--(1,0);
\draw (20:2)--(1,0);
\node at (-24:2.3) {$\pa\z_B$};
\node at (20:2.3) {$\pa\z_B$};	
\node at (111:-1.4) {(b)};	
\end{scope}
\end{tikzpicture}
\hspace{.3in}
\begin{tikzpicture}[line width=1 pt, scale=1]
\begin{scope}[shift={(1,0)}]
\draw (-140:1)--(0,0);
\draw (140:1)--(0,0);
\draw[vector] (.52,0) circle (.47cm);	

\node at (-140:1.3) {$\z_B$};
\node at (140:1.3) {$\z_B$};

\draw (-21:2)--(1,0);
\draw (20:2)--(1,0);
\node at (-23:2.25) {$\z_B$};
\node at (22:2.3) {$\z_B$};	
\node at (113:-1.4) {(c)};	
\end{scope}
\end{tikzpicture}
\nn\\
&&\hspace{.5in}\mbox{ Figure 4: diagrams with four external scalar lines}  \nn
\eea
The computation for the diagrams in Fig.4 (a) in the flat background\footnote{The computation in the AdS background can be found in Appendix B. Here we give more steps (than with the other diagrams) for the purpose of comparison with the AdS calculation.} goes as follows:
\bea
&&\hspace{1.3in}\fr{i^2}2<\int V_{hh\pa\z_B\pa\z_B}(x)\int V_{hh\pa\z_B\pa\z_B}(x')>\nn\\
&&\hspace{-.6in}=-\fr18 \int_u \sqrt{-g_0}\int_v \sqrt{-g_0}\Big<\Big(h^{\m\r}h_{\r}^\n-\fr14 {g}_0^{\m\n}h_{\r\s}h^{\r\s} \Big)
\Big(h^{\m'\r'}h_{\r'}^{\n'}-\fr14 { g}_0^{\m'\n'}h_{\r'\s'}h^{\r'\s'} \Big)\pa_\m\z_B \pa_\n\z_B\; \pa_{\m'}\z_B \pa_{\n'}\z_B \Big>
\nn\\
\eea
where $g_0^{\m\n}=\eta^{\m\n}$.
This can be written as
\bea
=-\fr18  \int_u \sqrt{-g_0}\int_v \sqrt{-g_0}\;\pa_\m\z_B \pa_\n\z_B\; \pa_{\m'}\z_B \pa_{\n'}\z_B \nn
\eea
\[
(g_0^{\k_1\m}g_0^{\k_4\n}g_0^{\k_2\k_3}-\fr14g_0^{\m\n}g_0^{\k_1\k_3}g_0^{\k_2\k_4} ) 
(g_0^{\k_1'\m'}g_0^{\k_4'\n'}g_0^{\k_2'\k_3'}-\fr14g_0^{\m'\n'}g_0^{\k_1'\k_3'}g_0^{\k_2'\k_4'} ) 
\Big<h_{\k_1\k_2} h_{\k_3\k_4}\;h_{\k_1'\k_2'} h_{\k_3'\k_4'} \Big>
\]
\[
\hspace{-.3in}=-\fr18  \int_u \sqrt{-g_0}\int_v \sqrt{-g_0}\;\pa_\m\z_B \pa_\n\z_B\; \pa_{\m'}\z_B \pa_{\n'}\z_B\;\D(u,v)\D(u,v) 
\Big(g_0^{\k_1\m}g_0^{\k_4\n}g_0^{\k_2\k_3}-\fr14g_0^{\m\n}g_0^{\k_1\k_3}g_0^{\k_2\k_4} \Big) 
\]
\bea
\Big(g_0^{\k_1'\m'}g_0^{\k_4'\n'}g_0^{\k_2'\k_3'}-\fr14g_0^{\m'\n'}g_0^{\k_1'\k_3'}g_0^{\k_2'\k_4'} \Big) (P_{\k_1\k_2\k_1'\k_2'}P_{\k_3\k_4\k_3'\k_4'}+P_{\k_1\k_2\k_3'\k_4'}P_{\k_3\k_4\k_1'\k_2'}) \nn\\ \la{4pazq}
\eea 
Noting that
\bea
&&\hspace{-1in} (g_0^{\k_1\m}g_0^{\k_4\n}g_0^{\k_2\k_3}-\fr14g_0^{\m\n}g_0^{\k_1\k_3}g_0^{\k_2\k_4} ) (g_0^{\k_1'\m'}g_0^{\k_4'\n'}g_0^{\k_2'\k_3'}-\fr14g_0^{\m'\n'}g_0^{\k_1'\k_3'}g_0^{\k_2'\k_4'} ) 
(P_{\k_1\k_2\k_1'\k_2'}P_{\k_3\k_4\k_3'\k_4'}+P_{\k_1\k_2\k_3'\k_4'}P_{\k_3\k_4\k_1'\k_2'})\nn\\
&& \hspace{1in} = \Big(\fr1{16}\eta^{\m\n'}\eta^{\m'\n} +\fr1{16}\eta^{\m\n}\eta^{\m'\n'}-\fr5{16}\eta^{\m\m'}\eta^{\n\n'} \Big)
\eea
the equation \rf{4pazq} takes
\bea
&=&-\fr18  \int_u \sqrt{-g_0}\int_v \sqrt{-g_0}\;\pa_\m\z_B \pa_\n\z_B\; \pa_{\m'}\z_B \pa_{\n'}\z_B\;\D(u,v)\D(u,v) \nn\\
&& \Big(\fr1{16}\eta^{\m\n'}\eta^{\m'\n} +\fr1{16}\eta^{\m\n}\eta^{\m'\n'}-\fr5{16}\eta^{\m\m'}\eta^{\n\n'} \Big)
\la{4paz2q}
\eea
After going to the momentum space and evaluating the momentum loop integral, this yields
 \bea
 \D\cL={\k'}^{4}\fr3{16}\fr{\G(\ve)}{(4\pi)^2}  \; (\pa \z_B)^4
 \eea
 where $C_1$ is defined in \rf{Cdef}. 
 As one can check, this flat background result is identical to those obtained by employing the AdS propagator in Appendix B. The computation of the other diagrams in Fig. 4 goes as follows. The correlator for the diagram in Fig. 4(b) turns out to vanish due to the tensor structure:
\bea
&&\hspace{-.6in}\fr{i^2}2 2<\int V_{hh\pa\z_B\pa\z_B}\int V_{hh\z_B \z_B}>
=\fr1{16}  m^2 \int\int \pa_\m\z_B \pa_\n\z_B\; \z_B^2 \Big<\Big(h^{\m\r}h_{\r}^\n-\fr14 \eta^{\m\n}h_{\r\s}h^{\r\s} \Big)
h_{\r'\s'}h^{\r'\s'} \Big>\nn\\
&& =0
\eea
The correlator for Fig. 4(c) is given by
\bea
&& \fr{i^2}2  \int \int  < V_{hh\z_B \z_B} V_{hh\z_B \z_B}>  
=-\fr1{128}  m^4 \int\int  \z_B^2\; \z_B^2 \Big<h_{\r\s}h^{\r\s} \;h_{\r'\s'}h^{\r'\s'} \Big>   \nn\\
\eea
Evaluating this one gets, for the counter-term,
\bea
\D\cL=\fr9{64} {{\k'}^{4}} m^4  \;\z_B^4
\eea
\vspace{.3in}

So far we have evaluated the various one-loop counter-terms in the $\overline{MS}$ scheme to be specific. The scheme that we actually employ when deriving the quantum-corrected field equations in the next section is the generalized $\overline{MS}$ scheme. 
Eventually, the terms entering the 1PI action are made finite by subtracting the counter-terms from the corresponding divergences. Let us illustrate this procedure by taking Fig. 1 and 3 that have the same forms of the counter-terms.
In the generalized $\overline{MS}$ scheme the finite parts  that one gets after subtracting the counter-terms from the divergences can be written as
\bea
 e_2\Rt^2+e_3\Rt_{\m\n}\Rt^{\m\n}    \la{totalctr}
\eea
where the constants $e_2,e_3$ can be fixed once the renormalization scheme is chosen.

\section{Violation of Dirichlet condtion}


Before we start to examine the boundary condition let us check the one-loop renormalizability of our system so that the observation of violation does not get compromised, at least at one-loop, by the issue of the renormalizability.\footnote{Some of the results of this section were obtained in collaboration with B.-H. Lee and K.-W. Lee.} As a matter of fact the field redefinition required for establishing the renormalizability already indicates (or at least hint at), as we will now see, that  the tension between the Dirichlet boundary condition and the quantum corrections. (The existence of such a metric also indicates that the method of the series form of the solution and conclusion drawn from its analysis below will remain valid to the higher orders of the loops and $\k$.)
The divergences calculated in the previous section can be absorbed by the following field redefinition:
\bea
g_{\m\n}&\ra&  g_{\m\n}+\k^2\Big[l_0g_{\m\n}+l_1  g_{\m\n}R+l_2 R_{\m\n}+l_3 g_{\m\n}(\pa\z)^2+l_4 \pa_\m\z \pa_\n\z+l_5g_{\m\n}\z^2\nn\\
&&+\k^2\Big(l_6R\pa_\m\z \pa_\n \z+l_7R_{\m\n}(\pa\z)^2+l_8R_{\m\n}\z^2+l_9g_{\m\n}R(\pa\z)^2 \nn\\
&& \;\;\;\hspace{.5in}+  l_{10} g_{\m\n}R^{\a\b}  \pa_\a \z\pa_\b \z+l_{11}R^{\a}{}_{\m\n}{}^{\b}  \pa_\a \z\pa_\b \z+l_{12} g_{\m\n}R\z^2\Big) \nn\\
&&\hspace{-.8in}+\k^4\Big(l_{13} g_{\m\n}(\pa \z)^4+l_{14} \pa_\mu \z\pa_\n \z(\pa\z)^2+l_{15} g_{\m\n}\z^4
+l_{16} g_{\m\n}(\pa\z)^2\z^2 +l_{17} \pa_\m\z \pa_\n\z \z^2\Big)+\cdots\Big]\nn\\  \la{msg}
\eea
where the coefficients $l$'s are to be chosen in such a way as to absorb the divergences. One noteworthy feature of the renormalization procedure is the crucial role played by the cosmological constant term, as we will comment on below.
Once the classical solution \rf{clasol} is substituted, the shifted part of the metric above contains the terms without the $z$ power, and thus, the time-dependent boundary terms. This indicates that the new solution no longer satisfies the Dirichlet condition. (We will shortly confirm this more directly from another angle.) 

The fact that the divergences can be absorbed by the field redefinition \rf{msg} can be seen as follows. With the field redefinition the Einstein-Hilbert term and the cosmological constant term become
\bea
&&\fr1{\k^2}\sqrt{-g}R-\fr1{\k^2}2\L \sqrt{-g}\nn\\
&&\hspace{-.3in}\Rightarrow
\sqrt{-g}\Big[-2\Big(\fr{\L}{\k^2}+\fr{\d\L}{\k^2}-2\fr{\d\k \L}{\k^3}+2l_0\k^2 \L\Big)+ \Big(\fr1{\k^2}-2\fr{\d\k }{\k^3}+l_0-\L (4l_1+l_2)\Big)\,R\nn\\
&&+(l_1+\fr{l_2}2)R^2-l_2R_{\m\n}^2 {-} 4\L l_5 \z^2-\L(4l_3+l_4)(\pa\z)^2\nn\\
&&+\Big(l_5-\L \k^2(l_8+4 l_{12}) \Big)\z^2R+\Big(l_3+\fr{l_4}2-\L \k^2 (l_6+l_7+4l_9) \Big)R(\pa\z)^2\nn\\
&&-\Big(l_4+\L \k^2(4l_{10}-l_{11}) \Big)R^{\a\b}\pa_\a\z \pa_\b \z\nn\\
&&+\k^2(l_{10}-\fr12 l_{11}-l_6)RR^{\a\b}\pa_\a\z \pa_\b \z+\k^2( \fr{l_8}2+l_{12} )\z^2R^2-\k^2 l_{11} R^{\m\n}R^\a{}_{\m\n}{}^\b \pa_\a\z \pa_\b \z\nn\\
&&-\k^2 l_7 R_{\a\b}^2 (\pa \z)^2-\k^2 l_8 R_{\a\b}^2  \z^2+\k^2(\fr{l_6}2+\fr{l_7}2+l_9)R^2(\pa\z)^2\nn\\
&&-\L \k^4 (4l_{13}+l_{14})(\pa \z)^4-4\L \k^4 l_{15}\z^4-\L \k^4 (4l_{16}+l_{17})\z^2(\pa\z)^2\nn\\
&&+\k^4(l_{13}+\fr12 l_{14}) (\pa\z)^4R+\k^4l_{15} \z^4 R+\k^4(l_{16}+\fr12 l_{17})\z^2(\pa\z)^2 R  \nn\\
&&-\k^4 l_{14}R^{\m\n}\pa_\m\z \pa_\n\z (\pa \z)^2-\k^4l_{17} R^{\m\n}\z^2\pa_\m\z \pa_\n\z  \Big]
\la{eh}
\eea
The matter part shifts
\bea
&&\hspace{1in}-\fr12 \sqrt{-g}\;(\pa \z)^2-\fr12 \sqrt{-g}\;m^2 \z^2 \nn\\
&&\Rightarrow \sqrt{-g}\Big[ -\fr12 (1+\k^2 l_0)(\pa \z)^2-\fr12 m^2 (1+\k^2 l_0) \z^2
   \nn\\
 &&\quad+\fr12\k^2 l_2 R_{\a\b}\pa^\a\z \pa^\b \z -\k^2(\fr{l_1}2+\fr{l_2}4)R(\pa \z)^2 -\k^2 m^2({l_1}+\fr{l_2}4)R\z^2\nn\\
 &&\quad -m^2\k^2l_5 \z^4+\k^2(-\fr{l_3}2+\fr{l_4}4)(\pa\z)^4  -\k^2\Big(\fr{l_5}2+m^2(l_3+\fr{l_4}4)\Big)\z^2(\pa\z)^2\nn\\
&&\quad+(\fr{l_6}4-\fr{l_7}4-\fr{l_9}2)R(\pa\z)^4 
-\k^4\Big(\fr{l_8}4+\fr{l_{12}}2 +m^2(\fr{l_6}4+\fr{l_7}4+l_9)\Big)R\z^2(\pa\z)^2\nn\\
&&+\k^4(-\fr{l_{10}}2+\fr{l_{11}}4+\fr{l_7}2)R^{\a\b}(\pa\z)^2\pa_\a\z \pa_\b\z+\k^4(\fr{l_{8}}2-m^2{l_{10}}+m^2\fr{l_{11}}4)R^{\a\b}\z^2\pa_\a\z \pa_\b\z\nn\\
&&+\fr12\k^4 l_{11}R^\a{}_{\m\n}{}^\b \pa_\a\z \pa_\b\z \pa^\m\z \pa^\n\z-m^2\k^4(\fr{l_8}4+l_{12})R\z^4 \Big]
 \la{se}
\eea
The total shifted Lagrangian is given by the sum of \rf{eh} and \rf{se}:
\bea
&&\hspace{1.0in}\sqrt{-g}\Big(\fr1{\k^2}(R-2\L)  -\fr12 \;(\pa \z)^2-\fr12 \;m^2 \z^2\Big)\nn\\
&&\hspace{-.8in}\Rightarrow
\sqrt{-g}\Big[-2\Big(\fr{\L}{\k^2}+\fr{\d\L}{\k^2}-2\fr{\d\k \L}{\k^3}+2l_0\k^2 \L\Big)+ \Big(\fr1{\k^2}-2\fr{\d\k }{\k^3}+l_0-\L (4l_1+l_2)\Big)\,R
+(l_1+\fr{l_2}2)R^2-l_2R_{\m\n}^2\nn\\
&&+\Big(-\fr12 (1+\k^2 l_0)-\L(4l_3+l_4)\Big)(\pa \z)^2+\Big(-\fr12 m^2 (1+\k^2 l_0) { -} 4\L l_5 \Big) \z^2\nn\\
&&\hspace{-.5in}+\Big(l_5-\L \k^2(l_8+4 l_{12})-\k^2 m^2({l_1}+\fr{l_2}4) \Big)R\z^2+\Big(l_3+\fr{l_4}2-\L \k^2 (l_6+l_7+4l_9)-\k^2(\fr{l_1}2+\fr{l_2}4) \Big)R(\pa\z)^2  \nn\\
&&+\Big({ -}l_4 { -} \L \k^2(4l_{10}-l_{11}) +\fr12\k^2 l_2\Big)R^{\a\b}\pa_\a\z \pa_\b \z   \nn\\
&&+\Big(-\L \k^4 (4l_{13}+l_{14})+\k^2(-\fr{l_3}2+\fr{l_4}4)\Big)(\pa \z)^4+\Big(-4\L \k^4 l_{15}  -m^2\k^2l_5\Big) \z^4  \nn\\
&& -\k^2\Big(\fr{l_5}2+m^2(l_3+\fr{l_4}4) +\L \k^2 (4l_{16}+l_{17})\Big)\z^2(\pa\z)^2 \nn\\
&&+\k^4\Big(-m^2 (\fr{l_8}4+l_{12})+l_{15}  \Big)R\z^4+\k^4\Big(l_{13}+\fr12 l_{14} +(\fr{l_6}4-\fr{l_7}4-\fr{l_9}2)\Big)R(\pa\z)^4 \nn\\
&& +\k^4\Big( l_{16}+\fr12 l_{17}  -\fr{l_8}4-\fr{l_{12}}2 -m^2(\fr{l_6}4+\fr{l_7}4+l_9)\Big)R\z^2(\pa\z)^2\nn\\
&&+\k^4 \Big(-l_{14}-\fr{l_{10}}2+\fr{l_{11}}4+\fr{l_7}2\Big)R^{\a\b}(\pa\z)^2\pa_\a\z \pa_\b\z\nn\\
&&+\k^4\Big(-l_{17} +\fr{l_{8}}2-m^2{l_{10}}+m^2\fr{l_{11}}4\Big)R^{\a\b}\z^2\pa_\a\z \pa_\b\z 
+\fr12\k^4 l_{11}R^\a{}_{\m\n}{}^\b \pa_\a\z \pa_\b\z \pa^\m\z \pa^\n\z \nn\\
&&+\k^2(l_{10}-\fr12 l_{11}-l_6)RR^{\a\b}\pa_\a\z \pa_\b \z+\k^2( \fr{l_8}2+l_{12} )\z^2R^2-\k^2 l_{11} R^{\m\n}R^\a{}_{\m\n}{}^\b \pa_\a\z \pa_\b \z\nn\\
&&-\k^2 l_7 R_{\a\b}^2 (\pa \z)^2-\k^2 l_8 R_{\a\b}^2  \z^2+\k^2(\fr{l_6}2+\fr{l_7}2+l_9)R^2(\pa\z)^2 \Big]
 \la{ehse}
\eea
The subsequent renormalization procedure can be outlined as follows. (Since we do not pursue the precise evaluation of the various coefficients appearing above, they can be determined up to these conditions.) 
The shifted cosmological constant $\fr{\L}{\k^2}+\fr{\d\L}{\k^2}-2\fr{\d\k \L}{\k^3}+2l_0\k^2 \L$ and the coefficient of $R$ can be set to the values chosen according to one's renormalization conditions, and those conditions will determine $\d\L$ and $\d\k$ in terms of the other quantities. 
The coefficients of $R^2$ and $R_{\a\b}^2$ should be matched with the results of the evaluation of the diagrams in Fig. 3:
\bea
l_1+\fr{l_2}2=e_2 ,\quad -l_2=e_3 
\eea
which yields
\bea
l_1= e_2+\fr12e_3 \quad,\quad  l_2= -e_3
\eea	
These equations are valid up to the chosen renormalization scheme and conditions.
 Quick examination of the coefficients of $R(\pa\z)^2$, $R^{\a\b}\pa_\a\z \pa_\b \z$ and $R\z^2$ reveals
\bea
l_3=\cO(\k^2),\quad l_4=\cO(\k^2),\quad l_5=\cO(\k^2)
\eea
Specifically, the coefficients of $R(\pa\z)^2$, $R^{\a\b}\pa_\a\z \pa_\b \z$ should be set to zero and the coefficient of $R\z^2$ should match with the result of the diagram in Fig. 2 (b): 
\bea
&&\hspace{-.3in} l_4+\L \k^2(4l_{10}-l_{11}) +\fr12\k^2 l_2=0,\quad l_3+\fr{l_4}2-\L \k^2 (l_6+l_7+4l_9)-\k^2(\fr{l_1}2+\fr{l_2}4)=0\nn\\
&&\hspace{.2in}  l_5-\L \k^2(l_8+4 l_{12})-\k^2 m^2({l_1}+\fr{l_2}4) =- 2\k^2 \fr{\G(\ve)}{(4\pi)^2} \fr{9 { m^2}}{16}
\eea
which can be viewed as constraints among $l_6,...,l_{12}$.
The terms in the fifth and sixth lines in the right-hand side of \rf{ehse} should be matched with the diagrams in Fig. 4. The terms in the seventh to tenth lines requires evaluation of the diagrams such as given in the following figure:
\bea
&& \hspace{-.2in}
\begin{tikzpicture}[line width=1 pt, scale=1]
\draw (-140:1)--(0,0);
\draw (140:1)--(0,0);
\draw[vector] (.52,0) circle (.47cm);	
\draw[vector] (.485,0.5)--(.485,1.3);

\node at (-137:1.3) {$\pa\z_B$};
\node at (137:1.3) {$\pa\z_B$};

\begin{scope}[shift={(1,0)}]
\draw (-40:1)--(0,0);
\draw (40:1)--(0,0);
\node at (-38:1.3) {$\pa\z_B$};
\node at (38:1.3) {$\pa\z_B$};	
\node at (65:-1.2) {(a)};	
\end{scope}
\end{tikzpicture}
\hspace{.3in}
\begin{tikzpicture}[line width=1 pt, scale=1]
\begin{scope}[shift={(1,0)}]
\draw (-140:1)--(0,0);
\draw (140:1)--(0,0);
\draw[vector] (.52,0) circle (.47cm);	
\draw[vector] (.485,0.5)--(.485,1.3);

\node at (-140:1.2) {$\z_B$};
\node at (140:1.2) {$\z_B$};

\draw (-21:2)--(1,0);
\draw (20:2)--(1,0);
\node at (-24:2.3) {$\pa\z_B$};
\node at (20:2.3) {$\pa\z_B$};	
\node at (111:-1.4) {(b)};	
\end{scope}
\end{tikzpicture}
\hspace{.3in}
\begin{tikzpicture}[line width=1 pt, scale=1]
\begin{scope}[shift={(1,0)}]
\draw (-140:1)--(0,0);
\draw (140:1)--(0,0);
\draw[vector] (.52,0) circle (.47cm);	
\draw[vector] (.485,0.5)--(.485,1.3);

\node at (-140:1.3) {$\z_B$};
\node at (140:1.3) {$\z_B$};

\draw (-21:2)--(1,0);
\draw (20:2)--(1,0);
\node at (-23:2.25) {$\z_B$};
\node at (22:2.3) {$\z_B$};	
\node at (113:-1.4) {(c)};	
\end{scope}
\end{tikzpicture}
\nn\\
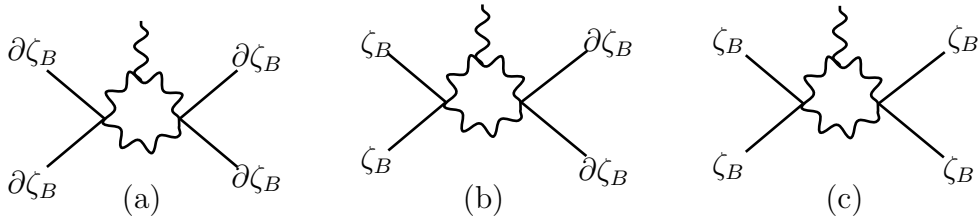
&&\hspace{.3in}\mbox{ Figure 5: diagrams of one metric and four scalar lines}  \nn
\eea
For the last two lines, the terms appearing in $(\cdots)$ in \rf{msg} become relevant and one should consider diagrams with two external graviton and two external scalar lines. We will not explicitly evaluate them in this work.\footnote{Additional terms such as $R_{\m\n}\z^4, g_{\m\n}R\z^4$ will have to be included in $(\cdots)$ in  the metric shift \rf{msg} once the analysis is extended to the higher order of $\k$ where those diagrams become relevant.}
One interesting feature of the analysis above is that the presence of the cosmological constant term is crucial for renormalizability. This feature will presumably remain valid in more complex systems such as a Einstein-Maxwell-scalar system.


\vspace{.3in}
With the outline of the renormalization procedure completed, let us now turn to the central issue of the violation of the Dirichlet boundary condition. 
It is now possible to write down the one-loop corrected action from the result obtained in section 3.
The 1PI action has the classical and quantum-correction parts: 
\bea
&&\hspace{.1in} S=\fr1{\k^2}\int d^4 x \sqrt{-g}\Big[R+\frac{6}{L^2}\Big] 
-\int d^4 x \sqrt{-g}\Big[\fr12(\pa_\mu \z)^2 +\fr12m^2\z^2\Big]\nn\\
&&\hspace{-.3in}+\fr1{\k^2}\int d^4 x \sqrt{-g}\Big[e_1{ \k^4} R\z^2+e_2 \k^2R^2+e_3\k^2 R_{\m\n}R^{\m\n}+e_4 { \k^6}(\pa\z)^4+e_5{ \k^6} \z^4\Big]  \la{qsact} \nn\\
\eea
where the tildes and subscripts `B's that appeared in the previous section have been removed. Consider the metric and scalar field equations that follow from \rf{qsact}:
\bea
&& R_{\m\n}-\fr12Rg_{\m\n}-\fr3{L^2}g_{\m\n}-\fr12g_{\m\n}\Big(-\fr12 { \k^2}(\pa_\mu \z)^2 -\fr12m^2 { \k^2}\z^2   \nn\\
&&\hspace{-.5in}+e_1 { \k^4}R\z^2 -2 e_1{ \k^4}\nabla^2\z^2 +e_2{ \k^2}R^2-4e_2{ \k^2}\nabla^2 R  +e_3{ \k^2} R_{\a\b}R^{\a\b}-e_3{ \k^2} \nabla^2 R +e_4{ \k^6} (\pa\z)^4+e_5{ \k^6} \z^4\Big)  \nn\\
&&\hspace{-.1in}-\fr12 { \k^2}\pa_\m \z \pa_\n\z+e_1{ \k^4}R_{\m\n}\z^2-e_1 { \k^4}\nabla_\m\nabla_\n \z^2+2e_2{ \k^2} RR_{\m\n}  -2e_2{ \k^2} \nabla_\m \nabla_\n R\nn\\
&& \hspace{-.1in} -2e_3{ \k^2} R_{\k_1\m\n\k_2} R^{\k_1\k_2}-e_3{ \k^2}\nabla_\m\nabla_\n R+e_3{ \k^2}\nabla^2 R_{\m\n}  +2e_4 { \k^6}\pa_\m \z \pa_\n\z (\pa\z)^2=0   \la{quanfe}
\eea
\bea
\nabla^2\z-m^2\z+2e_1\k^2 R\z-4e_4\k^4\Big[\nabla^2\z \, (\pa\z)^2+2\nabla_\a\z\,(\nabla^\a\nabla^\b\z)\,\nabla_\b\z\Big]+4e_5\k^4 \z^3=0\nn
\eea
One can write down the ansatze for the solution as the sum of the classical part plus quantum corrections. Consider the following quantum-corrected series form of the solution:
\bea
F(t,z)&=& F_0(t) +F_1(t) z+ F_2(t)z^2+F_3(t)z^3 + ...\nonumber\\
&+& (F_0^h(t) +F_1^h(t) z+ F_2^h(t)z^2+F_3^h(t)z^3 + ...) \nn\\
\Phi(t,z)&=&\frac{1}{z}+\Phi_0(t) +\Phi_1(t) z+ \Phi_2(t)z^2+\Phi_3(t)z^3 + ...\nonumber\\
&+& (\fr{\Phi_{-1}^h(t)}{z}+\Phi_0^h(t) +\Phi_1^h(t) z+ \Phi_2^h(t)z^2+\Phi_3^h(t)z^3 + ...)\nn\\
\z(t,z)&=&\z_0(t) +\z_1(t) z+ \z_2(t)z^2+\z_3(t)z^3 + ...\nonumber\\
&+&(\z_0^h(t) +\z_1^h(t) z+ \z_2^h(t)z^2+\z_3^h(t)z^3 + ...)   \la{1stans}
\eea
As in \cite{Murata:2010dx} one may first explore the case of $\Phi_0(t)=0$ with a similar condition imposed on its quantum counterpart:
\bea
\Phi_0(t)=0\quad,\quad \Phi_0^h(t)=0 \la{Diricon}
\eea
As discussed in the review section, one may utilize the residual gauge symmetry to fix the component $\Phi_0(t)$, and $\Phi_0(t)=0$ \cite{Murata:2010dx} is a gauge-fixing consistent with the Dirichlet boundary condition. At the quantum level, things become more subtle for the following reasons. Firstly, the coordinate change $\fr1{z}\ra \fr1{z}+g(t)$ amounts to introducing a different foliation. As well known and recently substantiated in the boundary theory context in \cite{Freidel:2016bxd,Donnelly:2016auv}, foliation matters at the quantum level: it introduces observer-dependent effects. Secondly, in general, gauge-fixing must be carried out in accordance with the physics one intends to study.  One may, for instance, consider physics associated with a different boundary condition by setting $\Phi_0(t)=f(t)$ where $f(t)$ denotes a non-vanishing fixed function. (If $\Phi_0(t)=0$ were a genuine gauge-fixing at the quantum level, which we doubt, it should be possible to prove the gauge-choice independence just as one proves the parameter independence of, e.g., the Feynman gauge in the Maxwell's theory.)\footnote{Another reason for the quantum-level subtlety for $\Phi_0(t)$-mode fixing is as follows. What has been shown in this work is that the one-loop (and higher-loop) quantum modes $\Phi_0^h$ is present. In a QFT, one fixes the gauge at the classical level and does not further adjust it at the quantum level. In other words, to make the quantum corrected solution obey the Dirichlet, one must remove the one-loop mode such as  $\Phi_0^h$ by the gauge transformation so that not only $\Phi_0$ but also  $\Phi_0^h$ gets removed. This is not something that one normally does: one does not adjust one's gauge-fixing at the quantum level; one just fixes it at the classical level and stays with it. (This of course means, as we have proposed, that the quantum corrected solution no longer obeys the Dirichlet.) Because of these reasons, we view, at the quantum level, $\Phi_0(t)=0$  as a part of the ansatze consistent with the Dirichlet boundary condition.}

The classical part of the analysis of \rf{quanfe} reproduces the results given in \rf{csol}:
\bea
&&\hspace{-.2in}\z _0=\Phi_0=F_1=0, \quad F_0=1, \quad  \Phi _1=-\frac{1}{8} \z _1^2,\quad F_2=-\frac{1}{4} \z _1^2,\quad
\Phi _2=-\frac{1}{6} \z _1 \z_2 \nn\\
&&\hspace{-.7in} \z _3=\frac{1}{2} \left(\fr12 \z _1{}^2 {\z}_1+2 \dot{\z} _2\right),\quad
\Phi _3=\fr1{96}\Big[ -\fr{11}4 \z _1{}^4 -8 \z _2^2 -12 {\z}_1 \dot{\z} _2\Big],\quad
\dot{F}_3=-\fr12\z_1\dot{\z}_2+\fr12 \z_1\ddot{\z}_2   \la{csol}   \la{cmodes} \nn\\ 
\eea
These results are parallel to those in \cite{Murata:2010dx} in which the case of a complex scalar field was analyzed. Interestingly, however, the quantum dynamics constrains $\z_1,\z_2$ as well so that
it now turns out that\footnote{This feature is also true for the complex scalar case.} 
\bea
\z_1(t)=0\quad ,\quad \z_2(t)=0 \la{psiz}
\eea  
With these, all the modes in \rf{cmodes} (exept $F_0$) vanish 
and the classical part of the metric becomes just that of AdS spacetime. This is intriguing: the classical black hole solution is not sustained. We will have more in the conclusion. For the quantum modes one gets
\bea
\z_0^h=0,\quad F_0^h=0,\quad
\Phi_{1}^h=0,\quad F_1^h=-2\dot{\Phi}_{-1}^h,\quad \Phi_2^h=0,\quad
F_2^h =0,\quad \z_3^h=\dot{\z}_2^h \nn\\
\eea
The presence of the modes $\Phi_{-1}^h$ implies that the quantum-corrected solution no longer satisfies the Dirichlet condition {(further remarks in the conclusion)}, and neither can it be interpreted as a Neumann-type boundary condition, as we will analyze below. The fact that a solution with nonvanishing $\Phi_{-1}^h$ exists implies that the boundary condition is unconventional yet-to-be discovered one. (It should presumably be a boundary condition dictated by removal of the high-order surface terms by the higher-order Gibbons-Hawking terms \cite{Gibbons:1976ue}.) It should reflect the nontrivial dynamics on the boundary and the information therein stored since the bulk modes depend on it.

Let us explore the case in which we do not impose \rf{Diricon}: 
\bea
\Phi_0(t)\neq 0\quad,\quad \Phi_0^h(t)\neq 0 \la{nDiricon}
\eea
Even with these, the outcome \rf{psiz} is not avoided, and the following relations are obtained: for the classical modes\footnote{These results were obtained with the help of a Mathematica package, diffgeo.m.}
\bea
&&F_0=1,\quad \z_0=0,\quad \Phi _1=-\frac{1}{8} \z _1^2=0,\quad F_1=2\Phi_0 \nn\\
&&\Phi _2=0,\quad F_2=\frac{1}{4} \left(4 F_0 \Phi _0{}^2-\zeta _1{}^2-8 \dot{\Phi} _0 \right)\nn\\
&& \dot{F}_3=0,\quad \z_3=0,\quad \Phi_3=0  ,\quad F_4=-F_3\Phi_0,\quad \z_4=0,\quad \Phi_4=0 \la{mrone}
\eea
and for quantum modes
\bea
&&\hspace{-.3in}\z_0^h=0,\quad F_0^h=0,\quad
\Phi_{1}^h=0,\quad F_1^h=2 \left(-\Phi _0(t) {\Phi}_{-1}^h-\dot{\Phi}_{-1}^{h}+{\Phi}_0^h\right),\nn\\
&&\Phi_2^h=0, \quad F_2^h=-2 \Big(-\Phi _{-1}^h \dot{\Phi} _0+\Phi _0^2 \Phi_{-1}^h-\Phi _0 \Phi_0^h+\dot{\Phi}_0^{h}\Big),
\nn\\
&&\hspace{-.5in}\quad \z_3^h=\Phi _0 \dot{\zeta}_1^{h}+\zeta_1^h \dot{\Phi} _0-\zeta_1^h \Phi _0^2-2
\zeta_2^h \Phi _0+\dot{\zeta}_2^{h},\quad { \dot{F}_3^h}=-3F_3\dot{\Phi}_{-1}^{h},\quad
\Phi_3^h=0\nn\\
&&\hspace{-.5in} F_4^h=F_3 \Phi _0 \Phi_{-1}^h-F_3 \Phi_0^h-F_3^h \Phi _0,\quad
\Phi_4^h=-3 {e_2} F_3 \Phi _0{}^2+3 {e_2} F_5-2 {e_3} F_3 \Phi _0{}^2+2 {e_3} F_5\nn\\
&&\z_4^h=4 {e_1} \zeta _4-\frac{1}{6} F_3 {\zeta_1^h}-3 \Phi _0{}^2 {\dot{\zeta}_1^h}-3 \Phi
_0(t) {\dot{\zeta}_2^h}-3 {\zeta_1^h} \Phi _0 \dot{\Phi} _0+\frac{4}{3} {\dot{\zeta}^h}_1 \dot{\Phi} _0 \nn\\
&&\hspace{.3in}+\frac{2}{3} \Phi _0 {\ddot{\zeta}_1^h}+\frac{2}{3} {\zeta_1^h} \ddot{\Phi} _0+2
{\zeta_1^h} \Phi _0{}^3+3 {\zeta_2^h} \Phi _0{}^2+\frac{2}{3} {\ddot{\zeta}_2^h}   \la{mrtwo}
\eea
Note that unlike their classical counterparts the modes $ \z_1^h$ and $ \z_2^h$ are not constrained.

The fact that the solution has non-vanishing time-dependent boundary components
\bea
\Phi_0(t)\neq 0, \quad \Phi_0^h(t)\neq 0,\quad\Phi_{-1}^h(t)\neq 0,
\eea 
implies that it no longer obeys the Dirichlet boundary condition at the quantum level. Let us examine whether or not it satisfies the Neumann boundary condition by any chance.
The variation of the Einstein-Hilbert term\footnote{To be complete, all of the surface terms arising from the quantum corrections terms in \rf{qsact} must be examined as well. Those terms would be cancelled by the generalized boundary terms in the extension of the standard practice. A related discussion can be found in \cite{Deruelle:2009zk,Teimouri:2016ulk}.} yields the following boundary terms:
\bea
\int d^4x \sqrt{-g}\; \nabla^{\a}\Big[  \nabla^{\b}\d g_{\a\b}- \nabla_{\a} (g^{\k_1\k_2}\d g_{\k_1\k_2}) \Big]
\eea
The standard practice is that one cancels these terms by adding the Gibbons-Hawking term \cite{Gibbons:1976ue}; one then imposes the Dirichlet boundary conditions (even though no boundary term remains since, for one thing, the equations of motion - which are partial differential equations - require boundary conditions). Suppose one does not add the Gibbons-Hawking term. In that case one must ensure that the boundary terms vanish by a different (i.e., non-Dirichlet) boundary condition.
The second term is the trace piece and therefore can be set apart \cite{Park:2015xoa}; let us focus on
\bea
\int d^4x \sqrt{-g}\; \nabla^{\a}  \nabla^{\b}\d g_{\a\b} 
\eea
Since the boundary is specified by a value of the $z$-coordinate the expression above yields
\bea
\int d^3y \sqrt{-g}\;   \nabla_{\b}\,\d g^{z\b} 
\eea
Note that
\bea
\nabla_{\b}\d g^{z\b} =\pa_{\b}\,\d g^{z\b}+\G^{z}_{\b\l}\d g^{\l\b}+\G^{\b}_{ \b\l}\d g^{z\l}
\eea
One should examine the behavior of the expression above in the region $z\ra 0$ after multiplying the determinant factor $\sqrt{-g}\sim \fr1{z^4}$.  Because the inverse metric and Christoffel symbols are $\sim z^2$ and $\sim \fr1{z}$ respectively, the boundary behavior of the solution cannot be interpreted as a Neumann-type boundary condition. This should be an indication that the boundary has its own full dynamics that cannot be handled simply by a boundary condition, which seems in line with the AdS/CFT dualities.

\section{Conclusion}

In this work we have looked into the issue of whether or not the classically consistent Dirichlet boundary condition could be extended to the quantum level. The issue requires analysis of a quantum-corrected time- and position- dependent solution. 
Given our goal, some of the technical complications could be avoided as mentioned in the introduction: those technicalities are needed to determine the coefficients $e$'s in \rf{qsact} more precisely including the genuine finite parts of the various Feynman diagrams. In other words the precise evaluation of the loop integrals require use of the actual propagator associated with the classical solution under consideration. However, the forms of the correction terms in the action \rf{qsact} are expected to arise rather generically and the precise determination of the coefficients was not necessary for demonstration of the quantum gravitational effects on the boundary. Furthermore, the renormalization scheme and conditions must be specified - a task we intend to postpone to the future works - in order to entirely fix the coefficients $e$'s.
We have obtained the 1PI action by first expanding around the usual AdS vacuum and taking the flat spacetime limit. (This procedure can also be viewed as taking the flat spacetime limit of the time- and position dependent solution itself.) We have then derived the field equations from the resulting action.  
The analysis has led to a clash between the Dirichlet boundary condition and quantum effects. One may force the solution to obey the Dirichlet boundary condition by removing such modes as $\Phi_0(t),\Phi_0^h(t),\Phi_{-1}^h(t)$ but then this seems to mean that the solution is a very special one possibly of ``measure-zero" in the solution space.

Several remarks are in order.
As mentioned above \rf{nDiricon}, the presence of the modes $\Phi_{-1}^h$ implies that the quantum-corrected solution no longer satisfies the Dirichlet condition. One may object this based on the reason that an appropriate boundary condition should be imposed on the quantum-corrected field equation. The present analysis reveals that the classical black hole solution will not remain as a solution (whereas the AdS solution does remain) with the Dirichlet condition imposed on the quantum corrected field equation. In other words, the Dirichlet condition can be taken as valid at the quantum level. (For the Dirichlet boundary condition at the quantum level, the higher-derivative gravity analogues of the Gibbons-Hawking terms would have to be added to cancel the surface terms.) But then the classical level black hole solution will not remain as a solution. {The quantum-corrected solution should presumably be a black hole solution. However, it will be substantially different, given the constraint \rf{psiz}, from the classical one.} One may view the ansatze \rf{1stans} as exploration of the possibility of a solution with a (yet to-be-understood) non-Dirichlet type boundary condition.

Although it may be possible to choose the $\overline{MS}$ scheme and remove the finite parts that would otherwise casuse the violation, the issue does not go away:  this just means that the bare action has been {\em chosen} in such a way as to avoid the issue. Moreover, the $\overline{MS}$ scheme is one of the many schemes that can be chosen.
The results in this work seem to imply that there exists a very direct link between the bulk and boundary dynamics in the sense that the bulk modes are given as functions of the boundary modes, \rf{mrone} and \rf{mrtwo}. The correlations between those modes seem to be in line with the pattern of the black hole information release observed in \cite{Park:2013rm}.

{One notable finding of the present work is that although the loop corrections are believed, at least naively, to be a small effect, the present analysis has revealed a possibility that the effect could be drastic after all. Perhaps this is not so surprising since Hawking radiaion, which is a quantum effect, leads to a drastic result, the evaporatioo of the black hole.}

\hspace{.2in}

There are several future directions worth pursuing. It will be interesting to numerically study various properties of the quantum corrected solution after fixing the coefficients $e$'s. But for this one must first examine whether or not there exists some type of boundary condition that remains valid at the quantum level. (See below for the related task.)
Secondly, as we have noted in the main body, the presence of the cosmological term was crucial for the renormalizability.
It could be that once the cosmological constant term is included, more complex systems may become at one-loop renormalizable in the conventional setup. For instance, it would be interesting to check  renormalizability for an Einstein-Maxwell-scalar system.
Another direction is Firewall-related.
With the one-loop renormalizability and the meaning of the field redefinition understood, it will be possible to directly tacke the Firewall issue \cite{Almheiri:2012rt} \cite{Braunstein:2009my,Braunstein:2014nwa} along the line of \cite{Park:2014mba}.
Still another direction is to seek a deeper understanding of the boundary dynamics. (See \cite{Freidel:2016bxd} for a recent development.) In order to come up with an action of the boundary physics, it is likely to be necessary to employ a new tactic such as the hypersurface reduction scheme \cite{Park:2013vpa}.




\newpage
\appendix

\renewcommand{\theequation}{A.\arabic{equation}}
 \setcounter{equation}{0}
\section{Conventions}

All the Greek indices are four-dimensional and all the Latin indices are three-dimensional
\bea
&&\a,\b,\g,...,\m,\n,\r...=0,1,2,3  \nn\\
&&a,b,c,...,m,n,r...=0,1,2
\eea
For the application of the BFM, we have introduced the unperturbed metric $g_{0\m\n}$, background field $\vf_{{}_B\m\n}$, and fluctuation $h_{\m\n}$, respectively:
\bea
g_{\m\n}\equiv  h_{\m\n}+\tilde{g}_{{}_B\m\n}\quad \mbox{where}\quad \tilde{g}_{{}_B\m\n}\equiv \vf_{{}_B\m\n}+g_{0\m\n} \la{gshift}
\eea
The graviton propagator is given by
\bea 
<h_{\m\n}(x_1)h_{\r\s}(x_2)>&=& P_{\m\n\r\s}\, \D(x_1-x_2) 
\eea
where, for the traceless propagator,
\bea
P_{\m\n\r\s}\equiv \fr12\Big(g_{0\m\r}g_{0\n\s}+g_{0\m\s}g_{0\n\r}
- \fr12g_{0\m\n}g_{0\r\s}\Big);
\eea
For a flat background,
\bea
\D(x_1-x_2)=\int \fr{d^4k}{(2\pi)^4}\fr{e^{ik\cdot (x_1-x_2)}}{i k^2}
\eea
In some places we have used
\bea
\k'^2 \equiv 2\k^2 
\eea

\renewcommand{\theequation}{B.\arabic{equation}}
\setcounter{equation}{0}
\section{Computations in AdS background}

As well known, the divergences are expected to come from the flat space limit. We confirm this by computing Fig.4 (a) by taking a flat limt\footnote{In the flat space limit the cosmological constant goes to zero. Although one naively expects to get a flat metric from the AdS metric, things are not entirely transparent in the Poincare coordinates. This aspect will presumably be more transparent in the static coordinates. However, one will face other complexities associated with using the angular variables. For example, one will have to use the special functions such as Bessel functions for the radial and angular directions whereas in the Poincare coordinates the plane waves can be employed for the $\vx$-directions.} of the intermediate AdS expression. As for the AdS computation, one should consider the subtlety associated with the lack of the conventional S-matrix in AdS spacetime prior to relating an offshell correlator to the corresponding S-matrix element. These issues were addressed, e.g., in \cite{Balasubramanian:1999ri,Giddings:1999qu}. (See e.g. \cite{Marolf:2012kh} for the case of dS.) For us, we keep things at the level of the offshell correlators.

The computation for the diagrams in Fig.4 (a) in the AdS goes as follows
The steps leading to \rf{4pazq} are the same as before (except that $g_{0\m\n}$ is now the metric of AdS).
For convenience we quote \rf{4pazq} here:
\[
\hspace{-.3in}=-\fr18  \int_u \sqrt{-g_0}\int_v \sqrt{-g_0}\;\pa_\m\z_B \pa_\n\z_B\; \pa_{\m'}\z_B \pa_{\n'}\z_B\;\D(u,v)\D(u,v) 
\Big(g_0^{\k_1\m}g_0^{\k_4\n}g_0^{\k_2\k_3}-\fr14g_0^{\m\n}g_0^{\k_1\k_3}g_0^{\k_2\k_4} \Big) 
\]
\bea
\Big(g_0^{\k_1'\m'}g_0^{\k_4'\n'}g_0^{\k_2'\k_3'}-\fr14g_0^{\m'\n'}g_0^{\k_1'\k_3'}g_0^{\k_2'\k_4'} \Big) (P_{\k_1\k_2\k_1'\k_2'}P_{\k_3\k_4\k_3'\k_4'}+P_{\k_1\k_2\k_3'\k_4'}P_{\k_3\k_4\k_1'\k_2'}) \nn\\ \la{4paz2}
\eea 
To examine the divergence coming from the flat spacetime limit let us set the $z$ coordinate to $\d$, a small positive number
\bea
z=\d \ra 0
\eea
Noting that
\bea
&&\hspace{-1in} (g_0^{\k_1\m}g_0^{\k_4\n}g_0^{\k_2\k_3}-\fr14g_0^{\m\n}g_0^{\k_1\k_3}g_0^{\k_2\k_4} ) (g_0^{\k_1'\m'}g_0^{\k_4'\n'}g_0^{\k_2'\k_3'}-\fr14g_0^{\m'\n'}g_0^{\k_1'\k_3'}g_0^{\k_2'\k_4'} ) 
(P_{\k_1\k_2\k_1'\k_2'}P_{\k_3\k_4\k_3'\k_4'}+P_{\k_1\k_2\k_3'\k_4'}P_{\k_3\k_4\k_1'\k_2'})\nn\\
&& \hspace{1in} =\d^4 \Big(\fr1{16}\eta^{\m\n'}\eta^{\m'\n} +\fr1{16}\eta^{\m\n}\eta^{\m'\n'}-\fr5{16}\eta^{\m\m'}\eta^{\n\n'} \Big)
\eea
where the factor $\d^4$ comes from the inverse metric factors.
The propagator becomes, in the limit $z\to \d$,
\bea
\D(x,x')|_{z\to \d} =\fr{2^{-\D}C_\D}{2\D-d}\;\xi^\D=\fr{2^{-\D}C_\D}{2\D-d}\;
\Big(\fr{2\d^2}{(\vec{x}-\vec{x}')^2}\Big)^\D
=\fr{C_\D}{2\D-d}\;
\fr{\d^{2\D}}{(\vec{x}-\vec{x}')^{2\D}}\nn\\
\eea
Noting that
\bea
d=3\quad,\quad \D=1 
\eea
the expression \rf{4paz2} becomes
\bea
\hspace{-.2in}= -\fr18  \int  \int \;  \fr{C_{1}^2}{(\vec{x}-\vec{x}')^{4}}  
\Big[\fr1{16} (\pa_\m\z_B \pa^\m \z_B) (\pa_{\m'}\z_B \pa^{\m'} \z_B)-\fr4{16} (\pa_\m\z_B \pa^{\m'} \z_B) (\pa_{\n}\z_B \pa^{\n'} \z_B)\Big]\nn\\
\eea
Let us introduce the Fourier transformations of the propagator and scalar fields\footnote{More rigorously, one should introduce a set of flat space coordinates $\tilde{x}^\m$ and set 
	\bea
	{ -\fr{C_1 }{2}}\fr{{ 2\d^2}}{(\vec{x}-\vec{x}')^{2}} & \rightarrow& \int \fr{d^{d+1}l}{(2\pi)^{d+1}}  \; \fr{e^{il\cdot (\tilde{x}-\tilde{x}')}}{il^2}  \nn\\
	\z_B(\vx,\d)&\rightarrow& \int d^d\vk_1 e^{i\vk_1\cdot \vec{\tilde{x}}} \zet(\vk_1,\d)  \la{fl}
	\eea
The coordinate $\tilde{x}^\m$ will be obtained by first going to the static coordinates and taking $\L\ra 0$ limit where $\tilde{x}^\m$  becomes the usual Cartesian coordinate.	
	}: 
\bea
{ -\fr{C_1 }{2}}\fr{{ 2\d^2}}{(\vec{x}-\vec{x}')^{2}} & \Rightarrow& \int \fr{d^{d+1}l}{(2\pi)^{d+1}}  \; \fr{e^{il\cdot (x-x')}}{il^2}  \nn\\
\z_B(\vx,\d)&\Rightarrow& \int d^d\vk_1 e^{i\vk_1\cdot \vx} \zet(\vk_1,\d)  \la{fl}
\eea
where $x-x'$ denotes $(0,\vx-\vx')$. Substituting these one gets
\bea
&&\hspace{-.5in}=  \fr1{8\d^4}  \ \int  \int  \; \int d^{d+1}l_1  \; \fr{e^{il_1\cdot (x-x')}}{il_1^2} \int d^{d+1}l_2   \; \fr{e^{il_2\cdot (x-x')}}{il_2^2}
\int d^d\vk_1  d^d\vk_2 d^d\vk_1' d^d\vk_2'   e^{i\vk_1\cdot \vx}e^{i\vk_2\cdot \vx} e^{i\vk_3\cdot \vx'} e^{i\vk_4\cdot \vx'}     \nn\\
&&\hspace{-.5in}\Big[\fr1{16}k_{1\m}k_2^\m k_{3\m'}k_4^{\m'}\zet(\vk_1,\d) \zet(\vk_2,\d)  \zet(\vk_3,\d)   \zet(\vk_4,\d)
-\fr4{16}k_{1\m} { k_{2\n}k_3^\m} k_4^{\n}\zet(\vk_1,\d) \zet(\vk_2,\d) \zet(\vk_3,\d)   \zet(\vk_4,\d)\Big]\nn\\
&&\hspace{-.5in}= \fr1{8\d^4}  {\clb L} \int d^d\vk_1  d^d\vk_2 d^d\vk_3 d^d\vk_4 \Big[\fr1{16}k_{1\m}k_2^\m k_{3\m'}k_4^{\m'}\zet(\vk_1,\d) \zet(\vk_2,\d) \zet(\vk_3,\d)   \zet(\vk_4,\d) \nn\\
&& -\fr4{16}k_{1\m} { k_{2\n}k_3^\m} k_4^{\n}\zet(\vk_1,\d) \zet(\vk_2,\d)  \zet(\vk_3,\d)   \zet(\vk_4,\d)\Big]   \d(\sum\vk)
\int d^{d+1}l     \;  \fr{1}{l^2 (l-k_3-k_4)^2}   
\la{4pazc}
\eea
The $d+1$-dimensional position space integral can formally be written as the $d$-dimensional integral:
\bea
\int d^dx \int dz \ra  L\int_\pa d^dx  
\eea	
where $\pa$ denotes the boundary and $L$ the size of the $z$-box. The parameter $L$ will disappear when converting the result back into the corresponding 4D expression at the end.
Evaluating the $l$-integral, the counter-term is given by 
\bea
\D\cL={ {\k'}^4}\fr3{16}\fr{\G(\ve)}{(4\pi)^2}  \; (\pa \z)^4
\eea
(The factor $\fr1{\d^4}$ in \rf{4pazc} will lead to $\sqrt{-g}$ in the counter-term action.)
This result is the same as the one obtained in the main body by applying the BFM entirely in the flat spacetime from the beginning.

\newpage

\end{document}